\documentclass[12pt,letterpaper]{JHEP3}
\usepackage{graphics}
\usepackage{epsfig}

\title{A covariant entropy conjecture on cosmological dynamical horizon}

\author{Song He\\
School of Physics, Peking
University, Beijing, 100871, China\\
 \email{she@pku.edu.cn}}

\author{Hongbao Zhang\footnote{Current address is Perimeter Institute for Theoretical Physics,
31 Caroline st. N., Waterloo, Ontario N2L 2Y5, Canada.}\\
Department of Astronomy, Beijing Normal University,
Beijing, 100875, China\\
\email{hbzhang@mail.bnu.edu.cn}}

\abstract{We here propose a covariant entropy conjecture on
cosmological dynamical horizon. After the formulation of our
conjecture, we test its validity in adiabatically expanding
universes with open, flat and closed spatial geometry, where our
conjecture can also be viewed as a cosmological version of the
generalized second law of thermodynamics in some sense.}


\begin{document}
\section{Introduction and motivation}
Recently Bousso has conjectured that in a spacetime satisfying
Einstein's equation with the dominant energy condition holding for
matter, the entropy flux $S$ through any null hypersurface generated
by geodesics with non-positive expansion emanating orthogonally from
some two-dimensional spacelike surface of area $A$ must
satisfy\cite{Bousso0,Bousso1,Bousso2,Bousso3}
\begin{equation}
S\leq \frac{A}{4}.\label{Bousso}
\end{equation}
Not only does this conjecture improve an earlier suggestion of
Fischler and Susskind on cosmic holography\cite{FS}, but also
reduces to the spacelike entropy bound whenever the latter is
expected to hold. Furthermore, this conjecture can be interpreted as
a statement of the so called holographic principle, which is
believed to be manifest in an underlying quantum theory of
gravity\cite{Hooft,Susskind}.

Later a generalized covariant entropy bound was suggested by
Flanagan, Marolf, and Wald\cite{Flanagan}. Namely, if one allows the
geodesics generating the null hypersurface to terminate at another
two-dimensional spacelike surface of area $A'$ before coming to a
caustic, boundary or singularity of spacetime, one can replace the
above conjecture Eq.(\ref{Bousso}) by
\begin{equation}
S\leq\frac{A-A'}{4}.\label{FMW}
\end{equation}
Obviously, this more general bound implies both the original Bousso
entropy bound and generalized second law of thermodynamics for any
process of black hole formation.

It is noteworthy that in highly dynamical spacetimes, since the
dynamical horizon, foliated by the apparent horizons, generally
divides the normal region from the trapped or antitrapped one, it
plays a subtle role in constructing the null hypersurface mentioned
above in such spacetimes as growing black holes and expanding
universes\cite{Bousso3}. In addition, as shown by Ashtekar and
Krishnan, there are series of intriguing properties related to the
dynamical horizon itself, such as the area balance
law\cite{AK0,AK1,AK2}. All of these motivate us to propose a
covariant entropy bound conjecture related to the dynamical horizon
in a direct way. In particular, as a first step, we here shall
suggest such a entropy bound on the dynamical horizon in the
cosmological context, which will be formulated in the next section.
In Section \ref{validity}, its validity is demonstrated in
adiabatically expanding universes. Conclusions and discussions are
presented in the end.
\section{Covariant entropy conjecture related to cosmological dynamical horizon}
Start from the FRW metric
\begin{equation}
ds^2=-dt^2+a^2(t)[\frac{dr^2}{1-kr^2}+r^2(d\theta^2+\sin^2\theta
d\phi^2)],
\end{equation}
which describes homogeneous and isotropic universes, including, to a
good degree of approximation, the portion we have seen of our own
universe. In terms of the conformal time $\eta$ and the comoving
coordinate, i.e.,
\begin{equation}
d\eta=\frac{dt}{a(t)}, d\chi=\frac{dr}{\sqrt{1-kr^2}},
\end{equation}
the FRW metric takes the form
\begin{equation}
ds^2=a^2(\eta)[-d\eta^2+d\chi^2+f^2(\chi)(d\theta^2+\sin^2\theta
d\phi^2)].
\end{equation}
Here $k=-1,0,1$ and $f(\chi)=\sinh\chi,\chi,\sin\chi$ correspond to
open, flat, and closed universes, respectively.

Next, to identify cosmological dynamical horizon, let us firstly
compute the initial expansion of the future directed null
congruences orthogonal to an arbitrary sphere characterized by some
value of $(\eta,\chi)$. Accordingly one finds\cite{Bousso3}
\begin{equation}
\theta_{\pm}=\frac{\dot{a}}{a}\pm\frac{f'}{f},\label{expansion}
\end{equation}
where the dot(prime) denotes differentiation with respect to
$\eta(\chi)$, and the sign $+(-)$ represents the null congruence is
directed at larger(smaller) values of $\chi$. Note that the first
term in Eq.(\ref{expansion}) is positive when the universe expands
and negative if it contracts. In addition, the second term is given
by $\coth\chi,\frac{1}{\chi},\cot\chi$ for open, flat, and closed
universes, respectively. Especially, this term diverges when
$\chi\rightarrow 0$, and it also diverges when $\chi\rightarrow \pi$
for a closed universe.

Now cosmological dynamical horizon is defined geometrically as a
three-dimensional hypersurface foliated by those spheres at which at
least there exists one orthogonal null congruence with vanishing
expansion. Thus cosmological dynamical horizon $\chi(\eta)$ can be
identified by solving the equation
\begin{equation}
\frac{\dot{a}}{a}=\pm\frac{f'}{f}.
\end{equation}
There is one solution for open and flat universes while for a closed
universe, there are generally two solutions, which are symmetrically
related to each other by $\chi_2(\eta)=\pi-\chi_1(\eta)$. Then a
cosmological version of our conjecture can be proposed as follows:
{\it{Let $A(\eta)$ be the area of cosmological dynamical horizon at
the conformal time $\eta$, then the entropy flux $S$ through
cosmological dynamical horizon between the conformal times $\eta$
and $\eta'$ must satisfy $S\leq\frac{|A(\eta)-A(\eta')|}{4}$ if the
dominant energy condition holds for matter. }}

Note that the description of our above conjecture is well defined
and obviously covariant. In the subsequent section, we shall test
its validity in adiabatically expanding universes.
\section{Covariant entropy conjecture tested by adiabatically expanding universes\label{validity}}
The matter content of FRW universes is most generally described by a
perfect fluid, with the energy momentum tensor
\begin{equation}
T_{ab}=a^2(\eta)\{\rho(\eta) (d\eta)_a(d\eta)_b+p(\eta)[(d\chi)_a
d(\chi)_b
+f^2(\chi)((d\theta)_a(d\theta)_b+\sin^2\theta(d\phi)_a(d\phi)_b)]\}.
\end{equation}
Later, we shall assume that the pressure $p$ and energy density
$\rho$ is related by a fixed equation of state
\begin{equation}
p=wp,
\end{equation}
where the constant $w$ is controlled within the range $-1\leq w\leq
1$ by the dominant energy condition. Furthermore, we shall restrict
ourselves in the case that FRW universes are in the expanding phase.
Thus, Einstein equation can be solved. As a result, the cosmological
scale factor is given by
\begin{equation}
a=f^q(\frac{\eta}{q}),
\end{equation}
where
\begin{equation}
q=\frac{2}{1+3w}
\end{equation}
is confined within the range $q\geq\frac{1}{2}$ or $q\leq -1$ due to
the dominant energy condition requirement. The corresponding
cosmological dynamical horizon is located at
\begin{equation}
\chi=\frac{\eta}{q}
\end{equation}
in all cases. An additional mirror horizon lies at
$\chi=\pi-\frac{\eta}{q}$ in the closed case\footnote{Due to this
space mirror symmetry, for a closed universe, we only need to focus
on cosmological dynamical horizon near the north pole $\chi=0$ in
the following discussions.}. Note that $\eta$ is positive for
$q\geq\frac{1}{2}$ and negative in the case of $q\leq -1$. Moreover,
$\frac{\eta}{q}<\frac{\pi}{2}$ is required in the closed case.

To proceed, we further assume that the evolution of matter in FRW
universes is adiabatical. Therefore the entropy current of matter
can be written as
\begin{equation}
s^a=\frac{s}{a^3}(\frac{\partial}{\partial t})^a,
\end{equation}
which implies the conservation of the entropy current, i.e.,
$\nabla_as^a$=0. Note that $s$ is actually the ordinary comoving
entropy density, constant in space and time.
\begin{figure}[htb!]
\hspace{-.1\textwidth} \vbox{\epsfxsize=1.2\textwidth
  \epsfbox{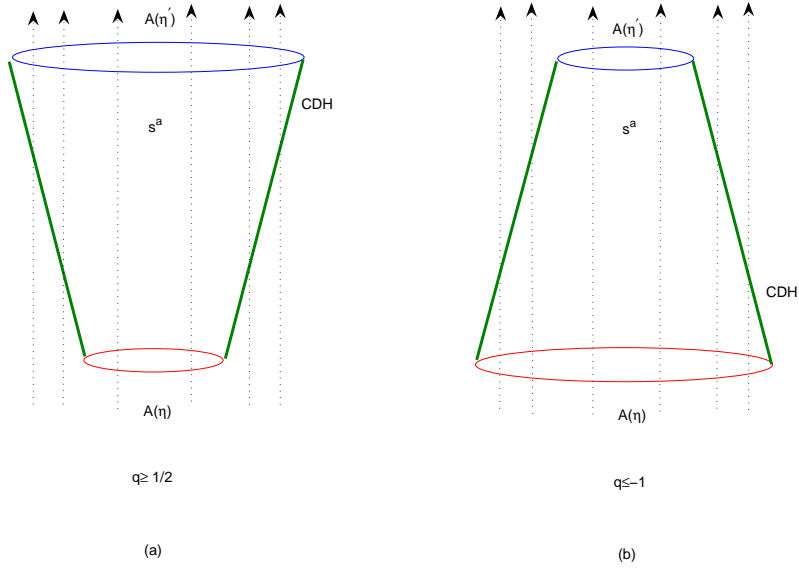}}
\caption {\small\sl (a)For $q\geq\frac{1}{2}$, the entropy current
flows across cosmological dynamical horizon from the normal region
to the antitrapped one. (b) For $q\leq -1$, the entropy current
flows across cosmological dynamical horizon from the antitrapped
region to the normal one.} \label{JHEP}
\end{figure}

We shall now check whether our conjecture is satisfied. However, as
shown in Figure \ref{JHEP}, there is an obvious difference between
$q\geq\frac{1}{2}$ and $q\leq -1$. Thus let $\eta'\geq\eta$, then by
the conservation of the entropy and Gauss theorem, our conjecture
can be reformulated as
\begin{equation}
\frac{A(\eta')}{4}-S(\eta')\geq \frac{A(\eta)}{4}-S(\eta)
\end{equation}
for $q\geq\frac{1}{2}$, and
\begin{equation}
\frac{A(\eta')}{4}+S(\eta')\geq \frac{A(\eta)}{4}+S(\eta)
\end{equation}
for $q\leq -1$\footnote{If we endow cosmological dynamical horizon
with a geometrical entropy of $\frac{A}{4}$ and take into account
the direction of the entropy current illustrated in Figure
\ref{JHEP}, it seems reasonable to assume that the total entropy of
universe is the geometrical entropy of cosmological dynamical
horizon plus that of matter in the antitrapped region for
$q\geq\frac{1}{2}$ while for $q\leq-1$ the total entropy is assumed
to be a sum of the geometrical entropy and that for matter in the
normal region. Then it is obvious that our conjecture can be
regarded as the generalized second law of thermodynamics for
expanding universes.}. Here we have invoked the fact that the area
\begin{equation}
A(\eta)=4\pi f^{2(q+1)}(\frac{\eta}{q})\label{area}
\end{equation}
is always an increasing function of $\eta$. In addition, $S(\eta)$
denotes the entropy flux through the normal region at the conformal
time $\eta$, given by
\begin{equation}
S(\eta)=4\pi s\int_0^\frac{\eta}{q}d\chi f^2(\chi).\label{entropy}
\end{equation}

Furthermore one finds our conjecture is equivalent to require that
\begin{equation}
H(\eta)\equiv \frac{1}{4}A(\eta)-S(\eta)\label{He}
\end{equation}
be an increasing function of $\eta$ for $q\geq\frac{1}{2}$.
Substituting Eq.(\ref{area}) and (\ref{entropy}) into Eq.(\ref{He}),
It follows
\begin{equation}
(q+1)f^{2q-1}(\frac{\eta}{q})-2s\geq 0.
\end{equation}
It is obvious that the LHS of inequality is an increasing function
of $\eta$ in all cases considered here. Whence we know that once the
inequality holds at some initial moment, it will continue to be
valid for all later moments. Note that at Planck epoch,
$\frac{\eta}{q}$ is of order one in all the cases. So the first term
in the LHS of the inequality is of the same order as $q+1$ which is
larger than $\frac{3}{2}$.  On the other hand, since the scale
factor is of the Planck size as well, the comoving entropy density
$s$ becomes the entropy per Planck volume which can not exceed one
quarter, as generally believed to be a generic feature for any
underlying theory of quantum gravity. Therefore in case of
$q\geq\frac{1}{2}$ our conjecture holds for open, flat and closed
expanding universes at Planck epoch and remains to be valid
henceforth.

Now let us turn to the case of  $q\leq -1$,  where our conjecture
means that the function
\begin{equation}
Z(\eta)\equiv \frac{1}{4}A(\eta)+S(\eta)
\end{equation}
is required as an increasing function of $\eta$, which follows
\begin{equation}
(q+1)f^{2q-1}(\frac{\eta}{q})+2s\leq 0.
\end{equation}
Similar to the case of $q\geq\frac{1}{2}$, the inequality will hold
for all the rest of the expansion epoch of the universe once it is
valid at an early time, say, Planck time.

At such an early time, the conformal time has decreased to
$\frac{\eta}{q}\sim O(1)$, thus the inequality holds if $s$ is
bounded by $-(\frac{q+1}{2})$ which is non-negative. In particular,
if $q=-1$, which describes a de-Sitter universe, the inequality
holds trivially due to $s=0$ for a cosmological constant.

Generally the bound for $s$ is $q$-dependent if we expect that our
conjecture holds at Planck time\footnote{Actually this is also the
case for $q\geq\frac{1}{2}$, although there is a reasonable
universal bound $\frac{3}{4}$ for $s$ there. It should be borne in
mind that the $q$-dependent bound for $s$ does not surprise us
because $q$ is a parameter associated with the matter and it is
definitely related to the entropy density of matter. In fact, in
\cite{Bousso3} the discussion is simplified by ignoring factors
containing $q$, and a more careful consideration will also lead to a
$q$-dependent bound for $s$ in test of Bousso entropy bound.}. If
$s$ turns out to be too large for our conjecture to hold at Planck
time, the validity of our conjecture will not be ruined, since it
will be valid at some later time and continue to hold henceforth.
The only difference is that the initial moment our conjecture
becomes valid is postponed due to a larger $s$. In other words, any
reasonable value of $s$ will yield a reasonable starting time for
our conjecture to hold, which is expected to be not too much later
than Planck time.

In all, our conjecture holds universally for open, flat and closed
universes which are adiabatically expanding with $q\geq\frac{1}{2}$
and $q\leq-1$, back to a very early time near Planck epoch(the exact
moment depends on $q$.). That is to say, once it is valid at some
early time, it continues to be.
\section{Conclusions and discussions}
We have proposed a covariant entropy conjecture on cosmological
dynamical horizon. Its validity has also been demonstrated in the
case of adiabatically expanding universes. In other words, if our
conjecture holds at some time in early universes, where classical
general relativity may be replaced by a quantum theory of gravity,
then it will hold for all the later times. The validity for our
conjecture at a time as early as near Planck time is also verified
under the assumption that $s$ is reasonably bounded. On the other
hand, although we restrict our discussion in expanding universes, it
is obvious that our conjecture also applies to collapsing universes
due to the time reversal symmetry.

All of these further encourage us to extend our conjecture to more
general contexts, especially to black hole dynamical horizons, whose
test is more difficult and challenging than that in cosmological
contexts, which will be reported elsewhere\cite{HZ1}. In addition,
it is highly desirable to provide some reasonable local conditions
sufficient for proof of our proposal and this work is also on
progress\cite{HZ2}.

\acknowledgments

We would like to thank Abhay Ashtekar for his stimulating series of
lectures on mathematical relativity and loop gravity at BNU. HZ is
also indebted to Xin Gao, Canbin Liang, Xiaoning Wu, and Wuhan Zhong
for helpful discussions on isolated and dynamical horizons. Work by
SH was supported by NSFC(nos.10235040 and 10421003). HZ was
supported in part by NSFC(no.10533010).

\end{document}